\documentclass[a4paper,aip,apl,preprint]{revtex4-1}

\usepackage{graphicx}
\usepackage{color}
\usepackage[dvipsnames]{xcolor}

\definecolor{bleu}{rgb}{0,0,1}
\definecolor{rouge}{rgb}{1,0,0}

\definecolor{aclock}{RGB}{162,18,42}
\definecolor{clock}{RGB}{17,143,64}

\newcommand{\Vu}{V\textcolor{bleu}{$\uparrow$}}
\newcommand{\Vd}{V\textcolor{rouge}{$\downarrow$}}

\begin{document}

\title[Single spin-torque vortex oscillator...]{Single spin-torque vortex oscillator using combined bottom-up approach and e-beam lithography}

\author{F.~Abreu~Araujo}\email{flavio.abreuaraujo@uclouvain.be (Author to whom correspondence should be addressed)}
\affiliation{Institute of Condensed Matter and Nanosciences, Universit\'{e} catholique de Louvain, Place Croix du Sud 1, 1348 Louvain-la-Neuve, Belgium}
\author{L.~Piraux}
\author{V.~A.~Antohe}
\affiliation{Institute of Condensed Matter and Nanosciences, Universit\'{e} catholique de Louvain, Place Croix du Sud 1, 1348 Louvain-la-Neuve, Belgium}
\author{V. Cros}
\affiliation{Unit\'{e} Mixte de Physique CNRS/Thales, 1 ave A. Fresnel, 91767 Palaiseau, and Univ Paris-Sud, 91405 Orsay, France}
\author{L.~Gence}
\affiliation{Institute of Condensed Matter and Nanosciences, Universit\'{e} catholique de Louvain, Place Croix du Sud 1, 1348 Louvain-la-Neuve, Belgium}

\begin{abstract}
A combined bottom-up assembly of electrodeposited nanowires and electron beam lithography technique has been developed to investigate the spin transfer torque and microwave emission on specially designed nanowires containing a single Co/Cu/Co pseudo spin valve. Microwave signals have been obtained even at zero magnetic field. Interestingly, high frequency vs. magnetic field tunability was demonstrated, in the range 0.4 - 2 MHz/Oe, depending on the orientation of the applied magnetic field relative to the magnetic layers of the pseudo spin valve. The frequency values and the emitted signal frequency as a function of the external magnetic field are in good quantitative agreement with the analytical vortex model as well as with micromagnetic simulations.
\end{abstract}

\pacs{85.75.-d, 75.47.De, 72.25.Ba}
\maketitle

The spin torque effect predicted by Slonczewski \cite{Slonczewski1996} and Berger \cite{Berger1996} has been demonstrated to be suitable for generating microwave signals from the injection of dc current through magnetic multilayer structures. In the last decade, several experimental works have shown the feasibility of the efficient generation of microwave oscillations of magnetization of single-domain ferromagnetic nanostructures \cite{Kiselev2003,Katine2008,Silva2008,Berkov2008} in the so-called spin-torque nano-oscillator (STNO). More recently, the vortex magnetic states have been considered to improve the signal coherence \cite{Pufall2007,Pribiag2007,Mistral2008,Ruotolo2009,Lehndorff2010,Dussaux2011,Yu2011,Locatelli2011,Darques2011,AbreuAraujo2012} and to obtain a larger power output \cite{Ruotolo2009,Locatelli2011,Dussaux2010} than previously with uniform mode oscillations. In the conditions of vortex based STNO, the STNO under study becomes the so called spin-torque vortex oscillator (STVO).

Various experimental approaches, materials and device configurations were explored for the implementation of suitable magnetic structures for microwave signal generation.  Still, the spectral purity and the power of the generated microwave signal should be greatly enhanced to reach the required values for future tunable microwave generators. To this aim, Dussaux \textit{et al.} \cite{Dussaux2010} have recently shown that MgO based magnetic tunnel junctions (MTJ) can generate a much higher level of power than metallic systems while keeping a reasonable spectral linewitdth.

\begin{figure}[h]
\centerline{\includegraphics[scale=.34]{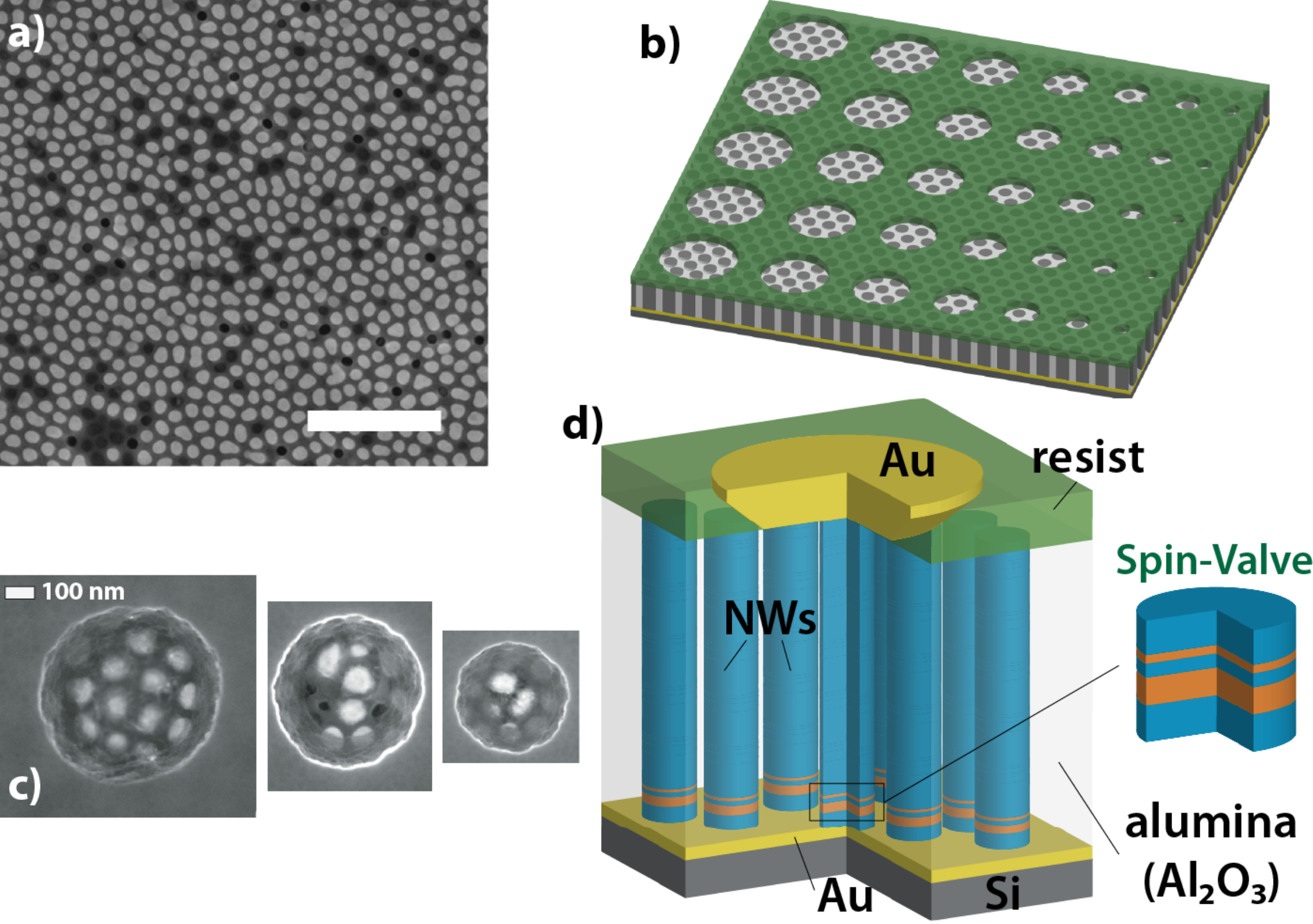}}
\caption{\label{fig:figure1a_1d} a) Top view scanning electron microscopy (SEM) image of the SV nanowire array embedded in the AAO template. b) Schematic showing a resist pattern from an assembly of circular apertures of different sizes. c) SEM images for 3 different openings showing the extremities of the nanowires at the surface of the template. d) Illustration of the completly structured device with single SVs located at the bottom of each NW.}
\end{figure}

It has been recently demonstrated \cite{Darques2011,AbreuAraujo2012} that STVOs based on nanowire (NW) spin valves (SVs) can be elaborated by electrodeposition, using anodic aluminum oxide (AAO) as templates \cite{Mourachkine2008,Blon2007,Wegrowe2002}. This bottom-up approach is a low cost and simple method to produce spin-transfer devices without complex processing. Amongst promising solutions envisaged to increase the generated power in STNOs (or STVOs) is a synchronization scheme with STNOs connected in parallel and/or in series as it is expected to increase the generated microwave power\cite{Georges2008}. Nevertheless, to the best of our knowledge, the synchronization of large arrays of STNOs is a technical issue that has not been solved so far. Indeed, experimental evidence for the coherent emission was recently reported using four STVOs \cite{Ruotolo2009}. In this respect, spin valve nanowires embedded in AAO templates present a high potential for obtaining an assembly of synchronized STNOs. Indeed, beyond its complementary metal-oxide semiconductor (CMOS) compatibility, excellent thermal and mechanical properties, tunable pore diameter and inter-pore distance, AAO templates supported on Si substrates exhibit a high pore density ($\sim 10^{10}$ pores/cm$^2$) which can be filled electrochemically to produce multilayered NWs, each NW consisting in $N$ SVs connected in series with $N \sim 1-10$. Till now, the approach based on atomic force microscope (AFM) nanoindentation of a thin insulating layer, was used to contact single NWs embedded in such AAO templates \cite{Fusil2005,Piraux2007}. In this work, we use a simple electron beam lithography (EBL) - based process suitable for a parallel connection of a well defined number of SV nanowires.
Unlike nanopillars lithographically patterned, our method is able to connect STNOs in parallel as well as in series. While the ability to connect electrically one to ten NWs in parallel has been demonstrated by means of AMR measurements on pure ferromagnetic NWs [Gence \textit{et al.}, to be published], here we report on the microwave emission from a single NW. Moreover each NW in the array contains a single (Co/Cu/Co) SV located at the bottom edge of the NW (see Fig. \ref{fig:figure1a_1d}(d)).

The sample fabrication process starts with the anodization of about a 1-$\mu$m-thick Al layer sputtered onto the Si/Nb(30nm)/Au(50nm)/Ti(6) substrate. The nanopores in the template are formed by complete anodic oxidation of the Al layer in 0.3 M oxalic acid solution at 2 $^\circ$C under a constant voltage of 60 V, followed by a chemical widening process of the pores in a 0.5M H$_2$SO$_4$ solution for about 2 hours at 40 $^\circ$C. The pore length and pore diameter at the bottom region are about 1.35 $\mu$m and 140 nm, respectively. The Au layer at the bottom of the pores serves as a working electrode for subsequent electrodeposition. NWs containing one spin valve Co(22nm)/Cu(12nm)/Co(6nm) have been electrochemically synthesized in a single bath using pulsed chronoamperometry. The composition of the aqueous electrolytic solution is as follows: CoSO$_4\cdot$7H$_2$O (1M), CuSO$_4\cdot$5H$_2$O (15mM) and H$_3$BO$_3$ (0.5M). Then, the pores were filled electrochemically with Cu (see Fig. \ref{fig:figure1a_1d}(a)). More details about the fabrication process and material characterization can be found elsewhere \cite{Piraux2007,Darques2006}.

Subsequently, electrical nanocontacts on a single or small number of wires was established using an EBL-based contacting technique. To this aim, the filled template was first thinned by mechanical polishing in such a way that the electrodeposited NWs end at the template surface and then thoroughly rinsed in order to obtain a very flat and clean top surface (see Fig. \ref{fig:figure1a_1d}(a)) required for further EBL processing.  The next step consists in covering the entire surface of the filled AAO template with a 300 nm thick silicon nitride (Si$_3$N$_4$) mask deposited by plasma enhanced chemical vapor deposition (PECVD) at low temperature $T=150$ $^\circ$C. A 200 nm thick polymethyl methacrylate (PMMA) layer is then spin-coated from a diluted solution 4\% in anisole (Microchemicals GmbH) and baked in an oven at 150 $^\circ$C  for 5 minutes. The electron beam exposure was done at 30 keV with an intensity of few tens of pA. The exposed pattern is developed in a mixture of methyl-isobutyl-ketone/isopropanol (IPA), rinsed in IPA and de-ionized water, and then blow dried with nitrogen. The pattern realized in PMMA is then transferred to the Si$_3$N$_4$ mask by reactive ion etching (RIE). By adjusting the exposure dose (see Fig. \ref{fig:figure1a_1d}(b,c)), a resolution better than 40 nm is achieved. Finally, electrical contact with the nanowires are made with 150 $\mu$m wide Au pads defined by EBL followed by Ti/Au metallization (see Fig. \ref{fig:figure1a_1d}(d)). By this method we were able to contact single wire of a dense NW forest.

\begin{figure}[h]
\centerline{\includegraphics[scale=0.8]{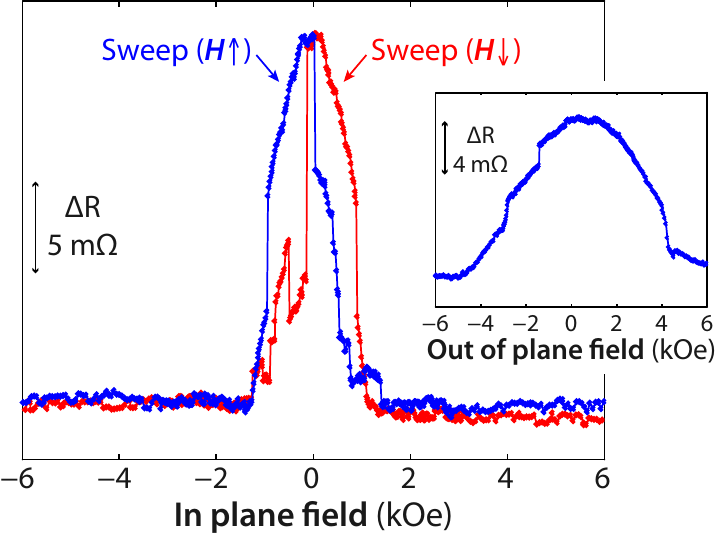}}
\caption{\label{fig:figure2} (Color online) GMR curves measured by contacting a single NW (made of a single Co(22nm)/Cu(12nm)/Co(6nm) SV) with the magnetic field applied in the  plane of the layers (IP field). The inset shows the GMR curve obtained by sweeping the external field along the wire axis (out of plane (OOP) field).}
\end{figure}

Electric transport measurements were then performed to characterize the resistance and magnetic configuration of the single NW.  Magnetoresistance curves obtained for a positive current of 4.0 mA are shown in Fig. \ref{fig:figure2}. The single NW resistance is about 18 $\Omega$. The GMR value is $\Delta R$ = 21 m$\Omega$ while sweeping the in-plane applied magnetic field and $\Delta R$ =12 m$\Omega$ in out of plane field (see inset of Fig. \ref{fig:figure2}). These results indicate that an antiparallel magnetization configuration can be more easily obtained when the field is oriented in the plane of the layers.  Also, as expected, the easy axis of the nanomagnet was set by the shape anisotropy to be perpendicular to the wire axis, i.e. in the plane of the layers.

\begin{figure}[h]
\centerline{\includegraphics[scale=.80]{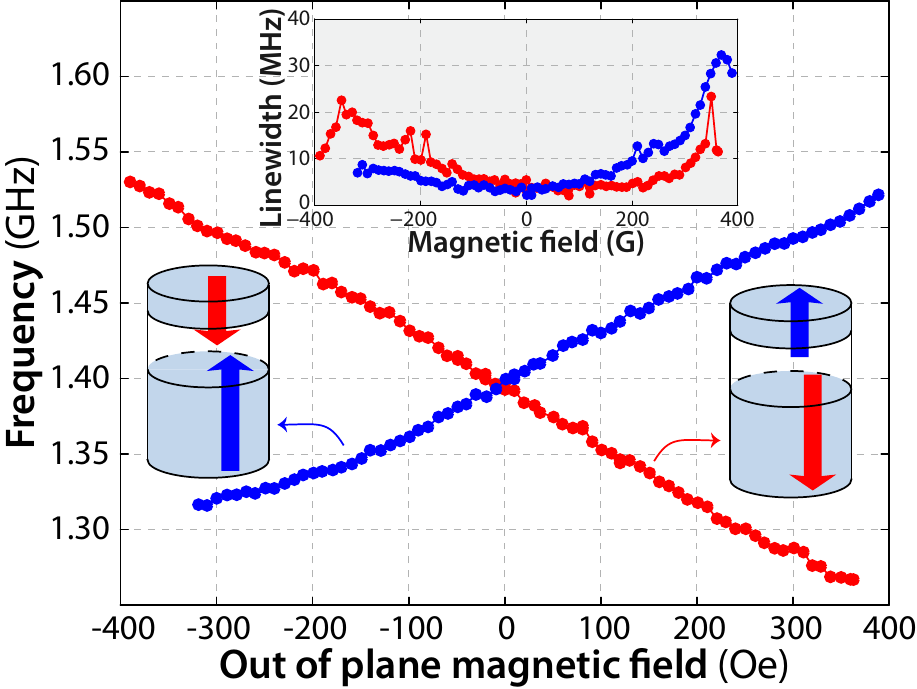}}
\caption{\label{fig:figure3} (Color online) Evolution of the emitted signal frequency as a function of the perpendicular applied magnetic field for the two magnetization configurations {\Vd/\Vu} (in blue) and {\Vu/\Vd} (in red). The signal was obtained by injecting a positive dc current of 6.0 mA. The inset shows the linewidth vs. magnetic field for both magnetization configurations.}
\end{figure}

\begin{figure}[h]
\centerline{\includegraphics[scale=.80]{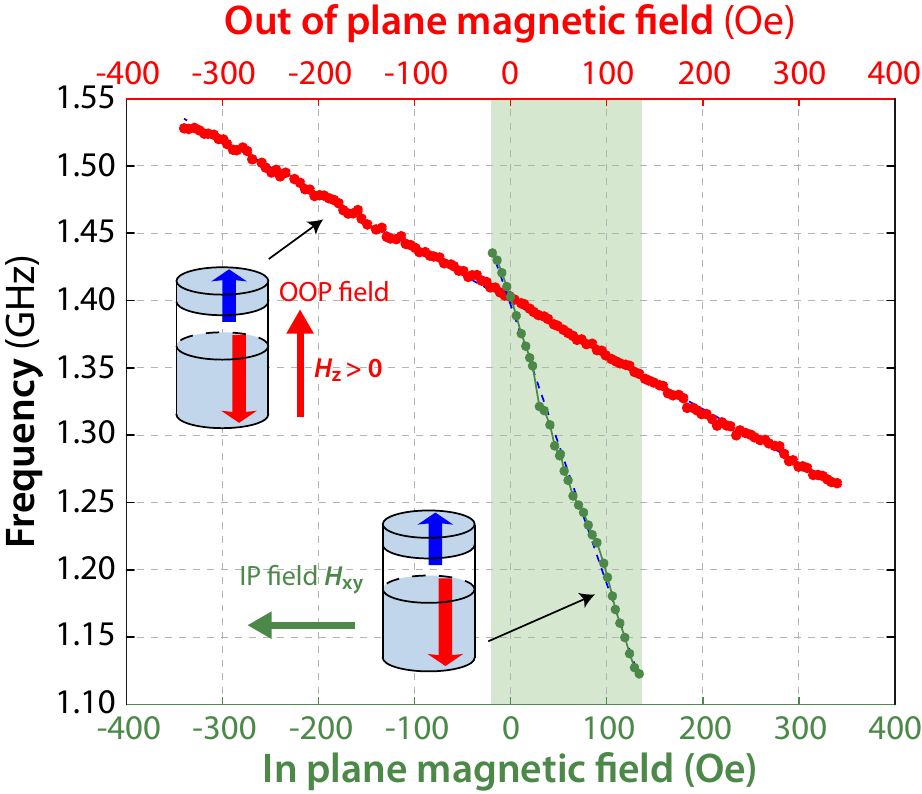}}
\caption{\label{fig:figure4} (Color online) Evolution of the emitted signal frequency as a function of the perpendicular (in red - top axis) and parallel (in green - bottom axis) applied magnetic field for the {\Vu/\Vd} magnetization configuration. The signal was obtained by injecting a positive dc current of 6.0 mA.}
\end{figure}

Microwave emission measurements are performed with a spectrum analyzer after a 42-dBm amplification of the output oscillating GMR signal. Selected experimental results of spin-transfer-driven vortex excitation obtained by measuring a single SV are displayed in figures \ref{fig:figure3} and \ref{fig:figure4}.

To stabilize the magnetic vortex state, a saturating perpendicular magnetic field is first applied to the sample before a decreasing ramp of magnetic field. Experimental evidence for the presence of vortex states in the Co layers has been demonstrated by Wong \textit{et al.} \cite{Wong2009} on the same type of electrodeposited multilayered NWs.

The results reported for the measured STVO are related to the two-vortex state with vortices of opposite polarity present in both the magnetic layers of the SV. The two possible magnetic configurations are {\Vu/\Vd} and {\Vd/\Vu}. {\Vu/\Vd} (resp. {\Vd/\Vu}) refers to a vortex with up (resp. down) polarity in the thin magnetic layer and a vortex with down (resp. up) polarity in the thick magnetic layer. The magnetic states responsible for the microwave emission are also consistent with micromagnetic simulations \cite{AbreuAraujo2012}.

By applying a perpendicular magnetic field and passing a dc current through the single SV, narrow peaks are observed in the recorded emission spectra. Figures \ref{fig:figure3} and \ref{fig:figure4} display the frequency vs magnetic field features for a positive injected dc current of $6.0$ mA.

Compared to our previous results \cite{AbreuAraujo2012}, the field range of emission has been increased. Indeed, the noise generated by the dipolar interaction between neighboring NWs in such a dense array is lowered as each NW contains only one SV (instead of 6 in the previous work). However, even if only one NW is contacted electrically, the dipolar interaction between adjacent NWs destabilizes the magnetic configuration as the magnitude of the magnetic field increases.

The sign of the slope of the frequency-field features (see Fig. \ref{fig:figure3}) is directly related to the relative alignment between the polarity of the vortex in the thick magnetic layer of the SV and the magnetic field direction. The experimental data are consistent with the remarkable properties of the magnetic configurations {\Vu/\Vd} and {\Vd/\Vu} as reported by De Loubens \textit{et al.} \cite{deLoubens2009} and Locatelli \textit{et al.} \cite{Locatelli2011}, and as shown in our previous work by means of micromagnetic simulations \cite{AbreuAraujo2012}. As the magnetization dynamics is essentially governed by the vortex located in the thick magnetic layer of the SV, the microwave signal frequency is compared with the analytically model. Extracting the vortex eigenfrequency  and the contribution of the DC current to the vortex core oscillation frequency (assuming $d f / d I = 40$ MHz/mA, as extracted from micromagnetic simulations) from refs \cite{Guslienko2002} and \cite{Choi2008} respectively, one obtains a vortex  gyrotropic frequency of about 1390 MHz. This value is nicely close to the experimental one, about 1400 MHz at zero field.

Thanks to the two-vortex states with opposite polarity the microwave emission appears even without any bias magnetic field and the frequency-field characteristics are linear. As shown in the inset of Fig. \ref{fig:figure3}, signal quality improvement is also obtained at zero field, i.e. the linewidth decreases. In contrast, when using a lithographically defined nanopillar SV the minimun of the linewidth is shifted and is obtained at a relatively large magnetic field, around 600 Oe as shown in ref\cite{Locatelli2011}.

Moreover, the relatively low linewidth observed in all the emission spectra is consistent with the vortex gyrotropic motion. Indeed, the measured linewidth was as low as 1.8 MHz for a positive dc current of 6.0 mA. The maximum power obtained is about 99 fW/mA$^2$. The results also corroborate the nucleation of a two-vortex state for positive currents as obtained in the micromagnetic study \cite{AbreuAraujo2012} thus giving rise to a better spectral quality of the signal (i.e., lower linewidth and larger peak height). One should notice that the current used for the measurements was limited to 6.0 mA to prevent from heating due to Joule effect, so that current densities as large as $\sim 6\cdot 10^{7}$ A/cm$^2$ were successfully injected on a single NW without deterioration.

Another remarkable property of our electrodeposited STVO is the large field-frequency tunability. It approximates 0.4 MHz/Oe for an out of plane field sweep (see red curve in Fig. \ref{fig:figure4}), but is even larger for an in plane field sweep. As shown in Fig. \ref{fig:figure4} (see green curve), the field-frequency tunability reaches 2 MHz/Oe. From these results, it appears that it is possible to produce a 300 MHz variation by sweeping the in plane magnetic field over a small range of 150 Oe.

As far as applications of STVOs are concerned, that kind of STVOs are very promising. Furthermore, the electron beam lithography is a versatile tool capable of connecting a single or a desired number of spin-torque vortex oscillators from the dense forest-like structure of nanowires embedded in nanoporous AAO templates integrated onto Si wafers with better compatibility to the standard CMOS processes.

F.A.A. acknowledges the Research Science Foundation of Belgium (FRS-FNRS) for financial support (FRIA grant).

%\bibliography{MyLib}{}

\end{document}